# Bioelectrical brain activity can predict prosocial behavior

Kunavin M.A.[1], Kozitsina (Babkina) T.S.[2,3*], Myagkov M.G.[4,5], Kozhevnikova I.S.[1,6], Pankov M.N.[1,6], Sokolova L.V.[1]

[1]Department of Human Biology and Biotechnical Systems, High School of Natural Sciences and Technologies, Northern (Arctic) Federal University, Arkhangelsk, Russian Federation

[2]Moscow Institute of Physics and Technology (National Research University), 9 Institutskiy per., Dolgoprudny, Moscow Region, 141701, Russian Federation

[3]Federal Research Center Computer Science and Control, Russian Academy of Sciences, Moscow, 119333, Russian Federation

[4]Laboratory of Experimental Methods in Cognitive and Social Sciences, Tomsk State University, Tomsk, Russian Federation

[5]Department of Political Science, University of Oregon, Eugene, OR, United States

[6]Institute of Medical and Biological Research, Northern (Arctic) Federal University, Arkhangelsk, Russian Federation

\* Correspondence:
Tatiana Kozitsina (Babkina)
tatyana.babkina@phystech.edu

## Abstract

Generally, people behave in social dilemmas such as proself and prosocial. However, inside social groups, people have a tendency to choose prosocial alternatives due to in-group favoritism. The bioelectrical activity of the human brain shows the differences between proself and prosocial exist even out of a socialized group. Moreover, a group socialization strengthens these differences. We used EEG System, "Neuron-Spectrum-4/EPM" (16 channels), to track the brain bioelectrical activity during decision making in laboratory experiments with the Prisoner's dilemma game and the short-term socialization stage. We compared the spatial distribution of the spectral density during the different experimental parts. The noncooperative decision was characterized by the increased values of spectral the beta rhythm in the orbital regions of prefrontal cortex. The cooperative choice, on the contrary, was accompanied by the theta-rhythm activation in the central cortex regions in both hemispheres and the high-frequency alpha rhythm in the medial regions of the prefrontal cortex. People who increased the cooperation level after the socialization stage was initially different from the ones who decreased the cooperation in terms of the bioelectrical activity. Well-socialized participants differed by increased values of spectral density of theta-diapason and decreased values of spectral density of beta-diapason in the middle part of frontal lobe. People who decreased the cooperation level after the socialization stage was characterized by decreased values of spectral density of alpha rhythm in the middle and posterior convex regions of both hemispheres.

## 1 Introduction

The actual issue of the modern cognitive research is the neurological involvement in the decision making processes, which requires the interdisciplinary approach of the use of social and neurobiological sciences (Kable & Glimcher, 2009; Schnuerch & Pfattheicher, 2018; Zhang & Gu, 2018). Nowadays, investigation of the collective decisions based on the analysis of the opponents' behavior became relevant and important (Fliessbach et al., 2007; Sanfey, 2007; Walter et al., 2004; Kozitsin et al., 2019, 2020). The collective games, which include behavioral economics and game theory compose the basis of the experimental models of such studies (De Quervain et al., 2004; King-Casas et al., 2005; Kuhnen & Knutson, 2005). The neurophysiological aspects of decision making are broadly being studied during the process of creating public goods (Chung, Yun, & Jeong, 2014). It is considered in the concept of a collective action problem (Gavrilets, 2015; Perry, Shrestha, Vose, & Gavrilets, 2018) and described in the example of altruistic punishment (Ciaramidaro et al., 2018; Mothes, Enge, & Strobel, 2016), free rider problem (Krajbich, Camerer, Ledyard, & Rangel, 2009), and prisoner's dilemma as well (Mahon & Canosa, 2012).

The prisoner's dilemma game is the most famous model to learn the decision making based on the cooperative and defective or noncooperative strategies. The psychophysiological aspects of these decisions include eye movements (Peshkovskaya et al., 2017), the analysis of functional magnetic resonance imaging (later fMRI), positron emission tomography (later PET) (Lukinova & Myagkov, 2016; Ramsøy, Skov, Macoveanu, Siebner, & Fosgaard, 2015; Sun et al., 2016), and evoked potential recording, and electroencephalography (later EEG) (Bell, Sasse, Oller, Czernochowski, & Mayr, 2016; Papageorgiou et al., 2013). EEG analysis allows characterizing the small changes in the areas of the human brain which are connected with cooperative or defective decisions. The bioelectrical signals from EEG are registered without time delays, that make it different from fMRI and PET. EEG sensors installation does not require special conditions, that make the participants' mobility possible during the experimental procedure.

The synchronization of the different EEG rhythms from the participants during the experimental procedure is studied through the simultaneous registration of EEG in several participants (Astolfi et al., 2010; De Vico Fallani et al., 2010; Dumas, Lachat, Martinerie, Nadel, & George, 2011; Hu et al., 2018). Such studies assume, that during the collective interaction between two individuals, their brain's bioelectrical activity should be considered as a single system (De Vico Fallani et al., 2010). Recent research in this area has led to the conclusion that visual contact between participants in the Prisoner's Dilemma game leads to the synchronization of their bioelectric activity in the temporal membrane region of the right hemisphere, which leads to the increased cooperation (Jahng, Kralik, Hwang, & Jeong, 2017). The activation of various areas of the prefrontal cortex in the decision making process is also accompanied by the decision making in the Prisoner's Dilemma game. Researchers indicate, that the left prefrontal cortex activation is basically connected with the logical evaluation of the performed decisions (Back, Carra, Chiaramonte, & Oliveira, 2009). In the studies with the use of low-frequency transcranial magnetic stimulation, the connection between cooperative behavior and the activation of dorsolateral regions of the left hemisphere is shown (Soutschek, Sauter, & Schubert, 2015). On the other side, competitive decisions stimulate the most activation in the orbitofrontal cortex and frontal pole (Astolfi et al., 2009).

One of the important issues in studying decision making is the mechanism facilitating the cooperation because this can play the leading role in the increase of collective action efficiency (Zak & Barraza, 2013). In neuroeconomics research, some results have already been obtained: the higher cooperation rate is correlated with the definite changes in the different brain areas (Fehr & Camerer, 2007; Gallese, Keysers, & Rizzolatti, 2004). Persons with a high level of empathy tend to make quickly cooperative decisions, that correlates with the activity of the superior temporal sulcus, and dorsolateral prefrontal cortex (Ramsøy et al., 2015). The ability to read the emotions and activities of others is connected with the system of mirror neurons activation underlying the front-limbic networks (Moll et al., 2006), which connects the medial prefrontal areas of the cortex (Amodio & Frith, 2006) with the subcortical nuclei of the striatum (Balleine, Delgado, & Hikosaka, 2007; Izuma, Saito, & Sadato, 2008).

In the experiments with the Prisoner's dilemma game, it is shown, that the socialization is an effective mechanism of the cooperative choice increase (Berkman et al. 2015). Socialization is the process to evoke prosocial behavior in a group of people and it refers to the social identity theory and small group paradigm (H. E. Tajfel, 1978; H. Tajfel & Turner, 1979). This can be due to the group selection, similarities among individuals, and good reputation (Riolo et al., 2001; Traulsen & Schuster, 2003; Nowak, 2006). Even the short-term socialization of strangers significantly increases the percentage of cooperative decisions (Babkina et al., 2016). The process of choosing the strategies in Prisoner's dilemma game differs between strangers and socialized participants. After socialization, the fixation and saccade frequency increased, but their duration decreased and on average the total duration of observing the task became smaller (Peshkovskaya et al., 2017). Changes after the socialization are observed also in the prefrontal cortex activation of the human brain, which was shown in fMRI study (Lukinova & Myagkov, 2016).

Thus, the paper is devoted to investigating the socialization effect on decision making in the Prisoner's dilemma game using EEG. The goal is to track the changes in the indicators of bioelectric activity of the brain, that arise in the process of the short-term socialization, as well as in decision making during the game before and after socialization. We hypothesized, that the socialization is accompanied by the specific changes in the brain activity. This finding can shed light on the neurobiology of the human sociality. The main instrumental technique was electroencephalography, which allowed us to record the bioelectric potentials of the brain directly during the game. The players are in the same room at all stages of the experiment, which is prohibited in fMRI.

## 2 Methods

Eight experiments were conducted. Participants (N = 96, 32 men and 64 women at the age of 18 to 36) were recruited as volunteers through the social network VKontakte (vk.com) and through the advertisements in the Northern (Arctic) Federal University (Arkhangelsk, Russia). Participants from the same experimental groups were unfamiliar with each other (from different departments, institutions, and courses). All the experiments were conducted in groups of 12 participants.

At the moment of the experiment all participants had no history of neurological or psychiatric disorders, were right-handed, and had normal or corrected vision. They were in a state of mental well-being, did not take drugs, that could affect the parameters of the brain functioning. We did not include in the study participants with a history of brain trauma or mental disorders.

The experiments were conducted with the standards of Good Clinical Practice and the principles of the Helsinki Declaration. The study protocol was approved by the Ethical Committee of the Northern (Arctic) Federal University (Arkhangelsk, Russia). Written informed consent was obtained from all participants after introduction of the experimental procedure. The data associated with this research are available at

https://dataverse.harvard.edu/privateurl.xhtml?token=a0beba2b-c6e3-46ea-a5cb-ff3c9ed6a52c.

We conducted all experiments in the class equipped with laptops connected to a single network. Every participant had an individual laptop and could not contact the other participants.

According to the experimental procedure, every participant earned some points which were converted into money at the end of the experiment (from 200 to 600 RUR that equivalents to the full lunch in cafeteria).

The experiments were based on the classical game theory model - Prisoner's dilemma game. A specialized tool to design and carry out group experiments in experimental economics, z-Tree developed at the University of Zurich, was used (Fischbacher, 2007). It consisted in three stages.

In the first stage (or before socialization), participants played the iterated Prisoner's dilemma game during the 22 periods or rounds. The iterated Prisoner's dilemma game consists of the number of iterated choices (periods) according with rules which will explain further. In our experimental design participants did not know about the number of repeats or periods. For them, the game was infinite. They played with the anonymous subject from the group and they knew that every period the opponent is changed randomly. The participants were able to move to the next period only after all 12 participants made their choices.

The Prisoner's dilemma game was based on the following rules. Each of two players in the Prisoner's dilemma game has two strategies: Cooperation (Up or Left) or Defection (Non-cooperation) (Down or Right). Two players gain the same points, R, for Cooperation and a smaller gain, P, for Defection. If one of the players cooperates and another defects, the cooperator gains a smaller reward, S, but the defector takes a larger reward, T. Thus, there is a ration between prizes $T>R>P>S$ (Table 1). Defection is more profitable than Cooperation in any partner's choice, but mutual Cooperation is more profitable for both than mutual Defection. The Nash equilibrium corresponds to mutual Defection (P, P), but the participants try to establish Cooperation (R, R) (Menshikov, I. (2017)).

In every period of the iterated Prisoner's Dilemma game participant should choose between Up and Down or Left and Right where Up and Left indicate Cooperation and Down and Right – Defection. So, based on the participant's choices during the periods of game it is possible to find the cooperation rate: the proportion of Cooperative choices from all choices.

After the first stage, the socialization stage was performed (the technique was from the behavioral studies) (Berkman, E.T. (2015), Peshkovskay, A. (2017), Peshkovskay, A. (2020)). First the participants memorized each other's names in a circle. Then the participants in reverse order shared specific thing or characteristic about themselves: hometown, hobbies, interests, and information about education. Then the experimenter asked about who would like to be a captain. And two captains were voluntary chosen. Other participants had to choose the captain whose team they wanted to join. In this way, two teams of six people with captains were formed based on the participants preferences. At the end of the socialization stage, in both groups participants were required to create a list of five common things unifying them. (favorite food, books, movies, etc.) and to decide the name of the team. This stage was approximately 15-20 minutes.

The last stage of the experiment (or after socialization) included 20 periods of the Prisoner's dilemma game inside the newly formed socialized groups of six. The rules of the games were the similar to the rules from the first stage. However, they again played with the anonymous subject and they knew that every period the opponent is changed randomly. But they also knew that the opponent is only from their group.

All the game results from the first and the third stages were automatically collected through the Z-Tree program.

In every experiment, two participants from twelve were randomly selected for electroencephalography (EEG) recording (N = 16, 8 men and 4). For EEG recording and analyzing we used Upgradeable EEG System with EP Capabilities "Neuron-Spectrum-4/EPM" (Neurosoft, Ivanovo, Russia). Standard electrode sites based on the international 10–20 system were used. The electrodes were placed unipolarity in the sixteen standard sites: frontal *(FP1, FP2, F3, F4, F7, F8)*, central *(C3, C4)*, temporal *(T3, T4, T5, T6)*, parietal *(P3, P4)*, occipital *(O1, O2)*. The separated ipsilateral reference electrodes were located on the ear lobes. The EEG recording was conducted when the participant was making decisions in front of the laptop screen during the first and the third stages of the game (during the second or socialization stage, participants with EEG recording were released from electrodes, so they were able to join all the procedures in this stage). The frequency of discrediting the EEG signal was 500 Hz. The EEG recording was carried out in the transmission band from 0.5 to 35 Hz.

Time-markers, corresponding to the moment, when experiment participant pressed chosen button in the game, were placed on the EEG records. The epochs of analysis were chosen relatively the exposed markers. Five-second pieces of the record before the markers were saved for analysis. Thus, we have obtained 42 epochs of analysis for each EEG record (22 before socialization and 20 after socialization). Saved epochs were cleared of artifacts using the independent components analysis (ICA). We used the Standardized Low Resolution Electromagnetic Tomography (sLORETA) to determine the brain sources. sLORETA is a functional imaging method based on certain electrophysiological and neuroanatomical constraints. The cortex was modeled as a collection of volume elements (voxels) in the digitized Montreal Neurological Institute (MNI) coordinates corrected to the Talairach coordinates.

We calculated EEG power spectral density for three basic rhythms: theta (4-8 Hz), alpha (8-13 Hz) and beta (13-30 Hz).

In order to identify the brain areas activated by the decisions making, the EEG activity of the non-cooperative choices was compared with the cooperative choices, making possible to compare brain activities associated with different performances. In order to identify the brain areas activated by the socialization, the EEG activity on the firth stage was compared with the EEG activity on the third stage.

Non-parametric statistical analysis of sLORETA (Statistical non-Parametric Mapping, SnPM) was performed for EEG segments in the respective spectral windows employing a log-F-ratio statistic for paired groups, with 5000 random permutations (i.e., bootstrapping) and levels of significance ($p < 0.05$) corrected for multiple comparisons and false positives. By this method, the SnPM bypasses the assumption of Gaussianity and reaches the highest possible statistical power.

# 3 Results

## 3.1 Behavior results

According to the rules of the iterated Prisoner's Dilemma game we calculated the cooperation rate for every participant – the ratio of the number of cooperative choices to all choices. On the first game stage before socialization the cooperation rate for all 96 study participants was 0.32±0.200 (Table 2). This parameter increased statistically significant and was 0.44±0.340 (Wilcoxon signed rank sum test, p = 0.010) after socialization. However, in the group of 16 participants who were registered with the parameters of bioelectric activity during the game, statistical significance was not achieved (Wilcoxon signed rank sum test, p = 0.236). In spite of this, the trend toward increase of cooperative decision number has remained. The mean share of cooperative decision before socialization was 0.38±0.269 and increased to 0.45±0.327 (Wilcoxon signed rank sum test, p = 0.236).

Based on this, we decided to analyze the socialization effect in two groups of study participants, differing by the cooperation rate at the different stages of study. One group included four participants whose cooperation rate after socialization increased from 0.41 to 0.77 (well-socialized). Second group included four participants whose cooperation rate after socialization decreased from 0.61 to 0.41 (no-socialized). This division on well-socialized and no-socialized was created based on the idea, that the socialization helps to promote sustainable cooperation among group members (Berkman et al. 2015). Thus, the effect of the socialization can be observed among the subjects who increased the cooperation rate.

## 3.2 Decision making

We have estimated the spectral density differences of EEG rhythms between cooperative and non-cooperative player choice at the first stage of analysis (Table 3). It has shown, that maximal intensity of low-frequency theta rhythm during cooperative choice has been registered in posterior part of frontal lobe above the superior frontal gyrus. 3D-visualization of theta-waves propagation allowed us to identify, that sources of this activity have been deep brain structures unassociated with cortex directly (Figure 1).

The choice of non-cooperative outcome of game round has been characterized by the increased values of spectral density of theta-diapason waves in temporal lobe of left hemisphere. Its maximum has been registered above the middle temporal gyrus.

Analysis of the spatial distribution alpha-rhythm spectral density allowed us to establish the choice of cooperative game outcome in comparison with non-cooperative choice accompanied by increased alpha-diapason waves in the frontal lobes of the cortex. Maximum of the spectral density has been registered above medial surface of prefrontal cortex of left hemisphere. Analysis of the spatial distribution of the beta rhythm spectral density has shown, that making non-cooperative decisions is characterized by an increase of activity in the orbital regions of prefrontal cortex. The results of analysis of EEG total power have been shown, that activity's main associative cortex areas decrease during the making a cooperative decision. Decreased EEG total power was registered in both frontal associative areas (medial and orbital cortex) and parietal associative areas.

## 3.3 Socialization Effect

At the second stage of analysis, we evaluated the socialization effect on the brain's bioelectrical activity of the study participants. Comparative analysis of EEG data recorded in the first experimental stage before socialization revealed the statistically significant differences of spectral density of all EEG-rhythms in participants from well-socialized and no-socialized groups (Table 4). Well-socialized participants differed in an increased level of the theta-rhythm spectral density in right hemisphere (Figure 2). Maximal differences were shown in the middle part of frontal lobe.

Initially low spectral density of the alpha rhythm was observed in the same participants group symmetrically in the both hemispheres. Maximal differences in comparison with the group of well-socialized participants were registered in the central regions of cortex. It should be noted, that the highest spectral density of alpha rhythm in the group of well-socialized participants was registered almost everywhere in the postcentral cortex regions with predominance in parietal areas of the both hemispheres.

Initially low spectral density of high-frequency activity was observed in the frontal regions of left hemisphere of the participants from well-socialized group who significantly increased the cooperation rate. Maximal between-group differences of beta-rhythm spectral density were registered in the middle part of frontal lobe.

Essentially similar results were shown when comparing the parameters of bioelectrical activity in participants from the both well-socialized and no-socialized groups after socialization (Table 5). Main tendencies were preserved in all analyzed frequency bands (Figure 3). The increased theta-activity in the frontal areas of right hemisphere continued to be observed in well-socialized participants of study. The highest values of spectral density of the alpha rhythm were noted in this group in the central regions of cortex exactly as before socialization. Decreased activity of beta rhythm was shown in the temporal areas of both hemispheres while maintaining reduced spectral power values in the frontal lobe of the left hemisphere.

# 4 Discussion

## 4.1 Decision making

In this paper we investigated the socialization effect on decision making in the Prisoner's dilemma game using EEG. We observed the changes in the indicators of bioelectric activity of the brain, that arise in the process of the short-term socialization, as well as in decision making during the game before and after socialization.

First of all, we compared the cooperative and non-cooperative choices in the Prisoners Dilemma game in terms of bioelectric activity of the brain. During cooperative choice, maximal intensity of low-frequency theta rhythm is registered above the superior frontal gyrus. Based on the result, we suggest, that activation of theta-rhythm during making of the cooperative decisions may be related to the functioning of limbic system and its anatomic formations such as amygdala, medial septum nucleus, and hippocampus. Critical role of these structures (septo-hippocampal system especially) for generation oscillations in theta-diapason has been demonstrated repeatedly in the studies of animals and humans (Buzsáki, G., 2002; Klausberger, T., Magill, P.J., Márton, L.F., 2003). For example, the limbic system is related functionally to regulation of internal organs, processes of memorization and activation memory traces as well as needs, motivations and emotions (LeDoux, J.E., 2000). The most important emotiogenic structures are considered to be the nuclei of amygdala complex and the cingulate gyrus, which are the subcortical center of emotional states regulation (Bush, G., Luu, P., Posner, M.I., 2000). Thereby, the choice of cooperation alternative is accompanied by more significant involvement of emotional sphere in decision making process and it is reflected the increase of theta-rhythm spectral density in precentral cortex regions.

About non-cooperative choice, we found, that it is characterized by the increased values of spectral density of theta-diapason waves in the temporal lobe of left hemisphere with maximum above the middle temporal gyrus. The theta-rhythm activation in these regions of the cortex is associated with actualization of memory traces and processes of reconsolidation carried out hippocampus. Its functional role in implementation of the working memory is determined by existence of extensive connections with the temporal and prefrontal cortex. On this basis, we propose, that making of the non-cooperative decisions during game is more based on the experience of the study participants and less on the analysis of emotional aspects of game situation.

Analysis of the spatial distribution alpha-rhythm spectral density reveals, that cooperative choice in comparison with non-cooperative is accompanied by increased alpha-diapason waves in the frontal lobes of the cortex with the maximum above medial surface of prefrontal cortex of left hemisphere in region of the anterior cingulate gyrus. According to modern concepts about the prefrontal cortex functions, its structures are integrated in a single block related to implementation of the executive functions such as processes of selective regulation of behavior, programming, goal-setting, planning and control over the targeted activities (Elliott, 2003; Diamond, 2013). The medial regions of prefrontal cortex, including the anterior parts of cingulate gyrus (24, 25 and 32 of the Brodmann areas) control behavior, assess adequacy of the performed or planned action to the goals and the expected results (Rushworth at al., 2004; Amiez et al., 2005; Buckley, Mansouri et al., 2009). The origin of alpha-oscillations is associated traditionally with inhibitory influences on the cortex from the thalamus and alpha rhythm is considered as «the rhythm of idleness». Thereby, the appearance of alpha waves in the medial regions of prefrontal cortex can indicate a decrease of rational control over making of cooperative decision.

Analysis of the spatial distribution of the beta rhythm spectral density shows, that making non-cooperative decisions is characterized by an increase of activity in the orbital regions of prefrontal cortex. Orbital parts of frontal lobe also take part in the implementation of executive functions, but unlike the medial prefrontal cortex, they regulate emotional and motivational aspects of behavior and social interaction (Ichihara-Takeda, Funahashi, 2007; Buckley, Mansouri et al., 2009). The orbital cortex has intensive connections with amygdala, basal ganglia and hypothalamus, included in limbic system and implements the voluntary regulation of emotional behavior (Croxson et al., 2005; Leh et al., 2010). Beta-rhythm is antagonist of alpha-waves and is formed as a result of the cortex neural networks activation. The increase of beta-rhythm spectral density in orbital cortex can indicate the enhancement of emotional control during making of non-cooperative decision. Thus, economically rational non-cooperative decision is more conscious and is making with tight voluntary control over performed action without emotions.

Finally, brain activity of main associative cortex areas decreases during making cooperative decision (in both frontal associative areas (medial and orbital cortex) and parietal associative areas). It should be noted, that successful and effective functional interaction of these cortical areas is an obligatory condition for formation of adequate programs of conscious activity on the basis of analysis of information on the environment. The effectiveness of functioning of front-parietal system is considered as one of the predictors of intelligence general level and ability for rational and logical thinking. Decreasing of this system activity during making of cooperative decision proves, that cooperation in the Prisoner's Dilemma game is more emotional and less rational decision.

## 4.2 Socialization Effect

According to the cooperation rate before and after socialization, we divided participants into two groups: well-socialized (who increased the cooperation after socialization) and no-socialized (who decreased the cooperation after socialization). Comparative analysis of EEG data reveals statistically significant differences of spectral density of all EEG-rhythms in participants from well-socialized and no-socialized groups. Well-socialized participants are differed in an increased level of the theta-rhythm spectral density in the right hemisphere (maximal differences were shown in the areas of middle frontal (10 BA)). The 10 Brodmann area, called frontal pole, is considered to be the most evolutionarily new cortex structure, which role is to coordinate various aspects of the organization of human behavior and to solve complex multicomponent tasks, that require the setting of intermediate goals subordinate to the final result (Burgess et al., 2007; Dumontheil et al., 2008). Differences in the representation of low-frequency EEG rhythms in these areas of individuals with the different socialization success may indicate a different effectiveness of their functioning. Probably, the highest activity of the frontal pole structures at the first experiment stage was a characteristic of the participants who subsequently showed a bad socialization effect and decreased the share of cooperative decisions in the game. And as previously noted, the functional role of lobe pole structures is the formation of multicomponent goals, the construction of action programs and control over their implementation. In the case of our experimental situation, a multicomponent goal could be winning the game (accumulating the maximum possible number of points) consisting of twenty individual rounds. The player had to think through the strategy of his behavior in each of the rounds, building a program of his actions while achieving intermediate results in such a way, that it would allow him to move towards the achievement of the final goal.

On the contrary, the middle and posterior convex regions of the hemispheres, including the parietal, occipital and temporal lobes are areas of projections of the main sensory systems and are engaged in analyzing and processing information about the environment, which also stores information in the long-term memory (Yeterian et al., 2012). In other words, the effective functioning of these structures is intended to produce an adequate assessment of the current situation. The increased spectral density of the alpha-rhythm in these areas of the cortex of well-socialized participants may indicate, that they are less prone to the in-depth analysis of their environment; make decisions and perform their actions more intuitively than based on the requirements of the situation.

Surprisingly, the socialization stage did not make any essential changes in the character of the bioelectric brain activity in both analyzed groups of the experiment participants. After the short-term socialization, the differences in the spatial distribution and activity of the main EEG rhythms appeared more clearly but did not undergo qualitative changes. It can be said, that the players who successfully passed the socialization stage and increased the share of cooperative decisions were initially set to the model of cooperative behavior. For this, there was a number of prerequisites for the functioning level of their brain. Persons ready for the socialization and cooperation relied less on careful analysis of the environment, made their decisions rather spontaneously and emotionally, their strategy for the game was not thought out and badly organized.

# 5 Conclusions

In this paper, we presented evidences, that a tendency to cooperative behavior can be detected via analysis of brain bioelectrical activity. First of all, we showed, that the cooperative and non-cooperative choices in the Prisoners Dilemma game differ in terms of EEG spectral density. Our results demonstrate, that defective decision is characterized by the increase of high-frequency beta activity in the orbital regions of prefrontal cortex (11 BA). It can reflect the involving of brain structures, that implement the voluntary regulation of emotional behavior. Thus, an economically rational, non-cooperative decision is more conscious and is made with tight voluntary control, and over performed action without emotions. The cooperative choice, on the contrary, is more instinctive or unconscious; it was accompanied by the increase of alpha activity in the anterior cingulate gyrus area (32 BA). The appearance of these waves in the medial regions of prefrontal cortex can indicate a decrease of rational control over making of cooperative decisions.

Our study also illustrates, that character of brain bioelectrical activity initially differs in participants who increased and decreased cooperation after the socialization stage. Well-socialized participants differ by increased values of spectral density of theta-diapason and decreased values of spectral density of beta-diapason in the middle frontal gyrus (10 BA). People who decrease the cooperation level after socialization are characterized by decreased values of spectral density of alpha rhythm in middle and posterior convex regions of both hemispheres. Surprisingly, the socialization stage did not make any essential changes in the character of the bioelectric brain activity in both analyzed groups of experiment participants. It can be said, that the players who successfully passed the socialization stage and increased the share of cooperative decisions were initially set to the model of cooperative behavior.

However, the method has the following limitation that should be taken into consideration. We used EEG system with sixteen electrodes because we conducted experiments in groups (not all the participants were selected for EEG recording). Moreover, the experimental procedure required that between the first and the third stages two participants with EEG were released from electrodes, but then we again equipped them with EEG. It was important to decrease the time when other participants wait for the participants with EEG. And we chose the minimal required number of electrodes to provide proper analysis. To avoid the problem with localization we described the broad brain regions. Nevertheless, these results give us more information about human sociality. A future investigation has the potential to develop theories about social behavior, and social identity theory (Ashforth, 1989).

**Acknowledgments**

We thank Rinat Yaminov for writing the programing code for experiments. This research was supported by The Tomsk State University competitiveness improvement program and by the grant in Russian Foundation for Basic Research (the grant number is 19-01-00296 A).

**Table 1 – Prisoner's Dilemma payoffs.**

| Payoffs | Left (Cooperation) | Right (Defection) |
|---|---|---|
| Up (Cooperation) | R, R | S, T |
| Down (Defection) | T, S | P, P |

For the experiments considered in this paper, the parameters of the Prisoner's dilemma were set as R = 5, P = 1, S = 0, T = 10 (10 > 5 > 1 > 0)

**Table 2 – Descriptive statistics of the behavior results.**

| Cases | Cooperation rate before socialization (Stage 1) | Cooperation rate after socialization (Stage 3) | p-value |
|---|---|---|---|
| All 96 participants | 0.32±0.200 | 0.44±0.340 | 0.01 |
| 16 participants with EEG recording | 0.38±0.269 | 0.45±0.327 | 0.236 |
| 4 participants with EEG recording – well-socialized | 0.41 | 0,77 | 0.011 |
| 4 participants with EEG recording – no-socialized | 0,61 | 0,41 | 0.031 |

**Table 3 –** Brain regions had shown the highest statistically significant differences in EEG activity between situations of making cooperative and non-cooperative decisions.

| Region | Gyrus | Brodmann areas | Hemisphere | Talairach coordinates | | | Log-F-ratio |
|---|---|---|---|---|---|---|---|
| | | | | X | Y | Z | |
| θ-rhythm | | | | | | | |

| Region | Gyrus | Brodmann areas | Hemisphere | X | Y | Z | Log-F-ratio |
|---|---|---|---|---|---|---|---|
| Temporal Lobe | Middle Temporal Gyrus | 21 | L | -64 | -44 | -6 | -0.422 |
| | Inferior Temporal Gyrus | 20 | L | -59 | -40 | -15 | -0.394 |
| | Superior Temporal Gyrus | 22 | L | -64 | -43 | 7 | -0.390 |
| Temporal and occipital Lobes | Fusiform Gyrus | 37 | L | -54 | -49 | -14 | -0.377 |
| Occipital Lobe | Middle Occipital Gyrus | 37 | L | -54 | -64 | -9 | -0.536 |
| Frontal Lobe | Superior Frontal Gyrus | 6 | L | -5 | -2 | 65 | 0.377 |
| | Medial Frontal Gyrus | 6 | L | -5 | -7 | 60 | 0.377 |
| Cingulate Cortex | Cingulate Gyrus | 24 | L | -5 | -7 | 46 | 0.361 |
| Frontal and parietal Lobes | Paracentral Lobule | 31 | L | -5 | -12 | 47 | 0.357 |
| α-rhythm | | | | | | | |
| Parietal Lobe | Inferior Parietal Lobule | 40 | L | -65 | -35 | 25 | -0,431 |
| Temporal Lobe | Superior Temporal Gyrus | 42 | L | -64 | -33 | 20 | -0.425 |
| Parietal Lobe | Postcentral Gyrus | 40 | L | -64 | -28 | 20 | -0.418 |
| Cingulate Cortex | Anterior Cingulate | 32 | L | -15 | 35 | 17 | 0.531 |
| Frontal Lobe | Medial Frontal Gyrus | 9 | L | -20 | 35 | 17 | 0.522 |
| β-rhythm | | | | | | | |
| Frontal Lobe | Superior Frontal Gyrus | 11 | L | -5 | 60 | -20 | -0.248 |
| | Orbital Gyrus | 11 | L | -5 | 43 | -23 | -0.247 |
| | Rectal Gyrus | 11 | L | -5 | 47 | -23 | -0.247 |
| | Medial Frontal Gyrus | 11 | L | -5 | 53 | -15 | -0.244 |
| Cingulate Cortex | Anterior Cingulate | 32 | L | -5 | 38 | -10 | -0.237 |

Log-F-ratio values more than 0.138 correspond statistically significant p < 0.05

**Table 4** – Brain regions had shown the highest statistically significant differences in EEG activity between groups of well-socialized and non-socialized gamers before socialization.

| Region | Gyrus | Brodmann areas | Hemisphere | Talairach coordinates | | | Log-F-ratio |
|---|---|---|---|---|---|---|---|
| | | | | X | Y | Z | |
| θ-rhythm | | | | | | | |

| Region | Gyrus | Brodmann areas | Hemisphere | X | Y | Z | Log-F-ratio |
|---|---|---|---|---|---|---|---|
| Frontal Lobe | Middle Frontal Gyrus | 10 | R | 35 | 39 | 12 | 1.07 |
| | Inferior Frontal Gyrus | 46 | R | 35 | 35 | 12 | 1.07 |
| | Superior Frontal Gyrus | 10 | R | 35 | 54 | 16 | 1.06 |
| | Precentral Gyrus | 44 | R | 59 | 15 | 13 | 1.04 |
| Insular cortex | Insula | 13 | R | 35 | 20 | 13 | 1.04 |
| α-rhythm | | | | | | | |
| Frontal and parietal Lobes | Paracentral Lobule | 6 | R | 5 | -31 | 66 | 2.01 |
| Parietal Lobe | Postcentral Gyrus | 4 | R | 10 | -31 | 66 | 2.01 |
| Frontal Lobe | Medial Frontal Gyrus | 6 | R | 5 | -26 | 66 | 2.01 |
| Cingulate Cortex | Cingulate Gyrus | 31 | L | -5 | -42 | 39 | 2.01 |
| Parietal Lobe | Precuneus | 7 | L | 0 | -41 | 48 | 2.00 |
| β-rhythm | | | | | | | |
| Frontal Lobe | Superior Frontal Gyrus | 10 | L | -20 | 63 | 6 | -0.89 |
| | Middle Frontal Gyrus | 10 | L | -25 | 59 | 11 | -0.87 |
| Cingulate Cortex | Anterior Cingulate | 32 | L | -15 | 39 | 12 | -0.86 |
| | Cingulate Gyrus | 32 | L | -10 | 30 | 26 | -0.78 |
| Temporal Lobe | Middle Temporal Gyrus | 21 | R | 64 | -34 | -7 | -0.75 |

Log-F-ratio values more than 0.497 correspond statistically significant p < 0.05

**Table 5** – Brain regions had shown the highest statistically significant differences in EEG activity between groups of well-socialized and non-socialized gamers after socialization.

| Region | Gyrus | Brodmann areas | Hemisphere | Talairach coordinates | | | Log-F-ratio |
| | | | | X | Y | Z | |
|---|---|---|---|---|---|---|---|
| θ-rhythm | | | | | | | |
| Frontal Lobe | Middle Frontal Gyrus | 46 | R | 40 | 44 | 7 | 1.08 |
| | Inferior Frontal Gyrus | 46 | R | 40 | 35 | 12 | 1.07 |
| | Superior Frontal Gyrus | 10 | R | 30 | 53 | -3 | 1.05 |
| | Precentral Gyrus | 44 | R | 59 | 20 | 8 | 1.03 |
| Insular cortex | Insula | 13 | R | 35 | 20 | 13 | 1.03 |
| α-rhythm | | | | | | | |
| Cingulate Cortex | Cingulate Gyrus | 31 | L | -15 | -42 | 25 | 1.51 |
| | Posterior Cingulate | 23 | L | -5 | -42 | 25 | 1.51 |
| Parietal Lobe | Precuneus | 31 | L | -10 | -47 | 30 | 1.50 |
| Temporal Lobe | Sub-Gyral | 31 | L | -20 | -47 | 35 | 1.48 |
| Frontal and parietal Lobes | Paracentral Lobule | 5 | L | -5 | -41 | 48 | 1.46 |
| β-rhythm | | | | | | | |
| Temporal Lobe | Middle Temporal Gyrus | 21 | R | 69 | -34 | -7 | -1.40 |
| | Inferior Temporal Gyrus | 20 | R | 59 | -35 | -15 | -1.36 |
| | Superior Temporal Gyrus | 22 | R | 59 | -34 | 6 | -1.33 |
| Temporal and occipital Lobes | Fusiform Gyrus | 37 | R | 50 | -44 | -6 | -1.32 |
| Cingulate Cortex | Anterior Cingulate | 32 | L | -15 | 35 | 17 | -1.27 |

Log-F-ratio values more than 0.496 correspond statistically significant $p < 0.05$

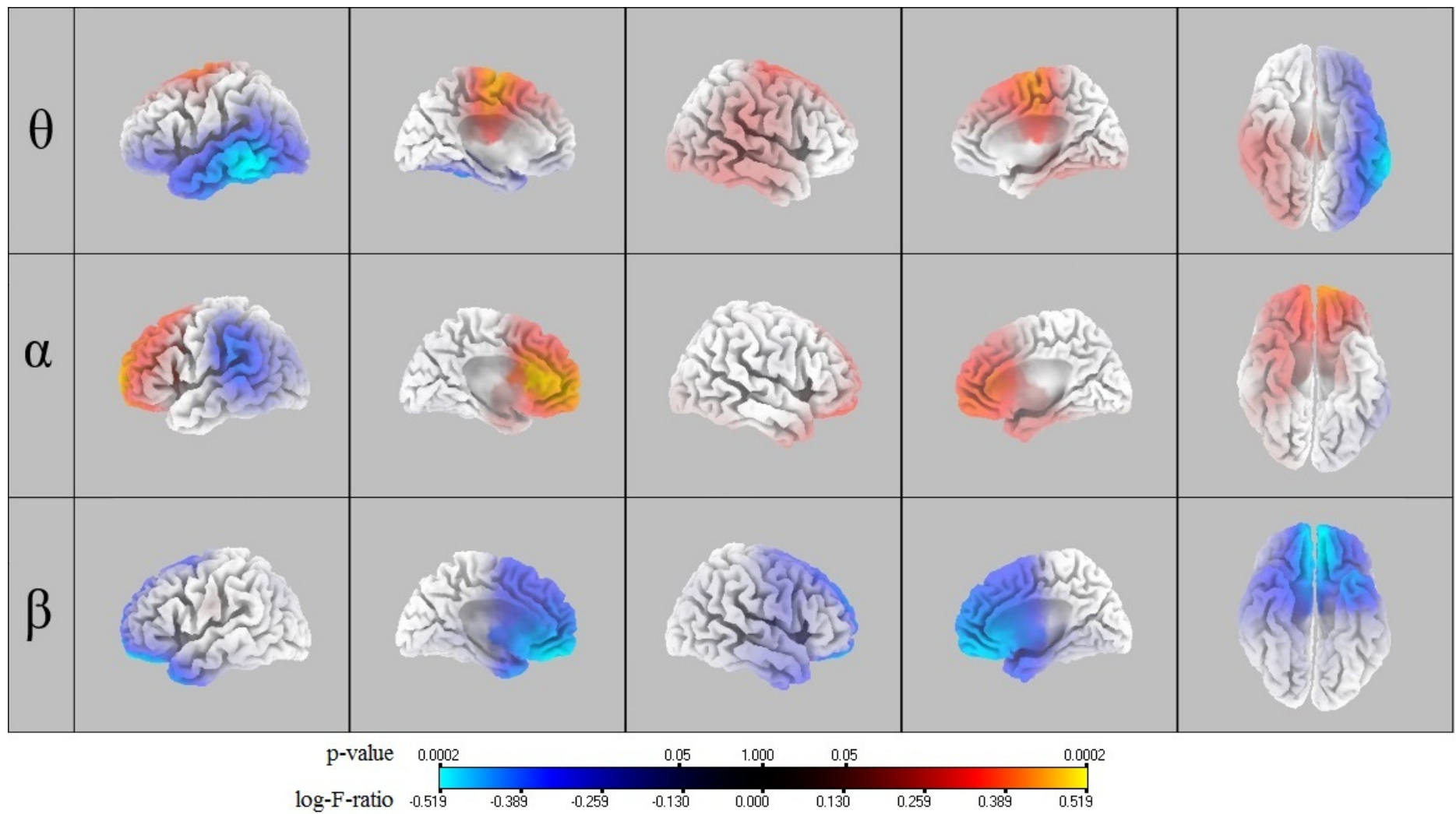

**Figure 1** – Cortical representation of the spectral density of main EEG rhythms during decision making in the Prisoner's Dilemma game. Red color is the higher activity during making of cooperative decisions. Blue color is the higher activity during making of non-cooperative decisions.

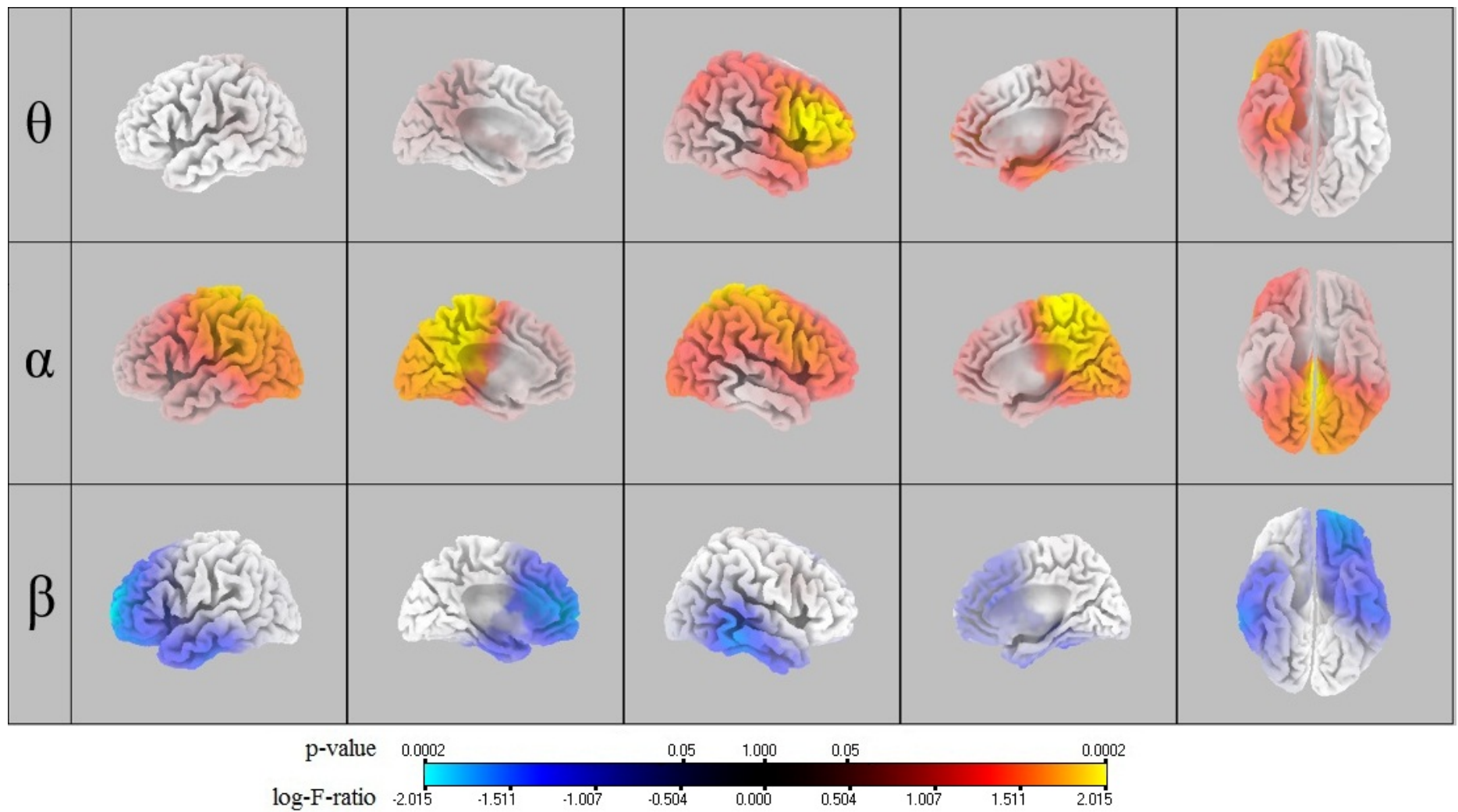

**Figure 2** – Between-group differences of cortical representation of the spectral density of main EEG rhythms before socialization. Red color is the higher activity in group of well-socialized gamers. Blue color is the higher activity in group of non-socialized gamers.

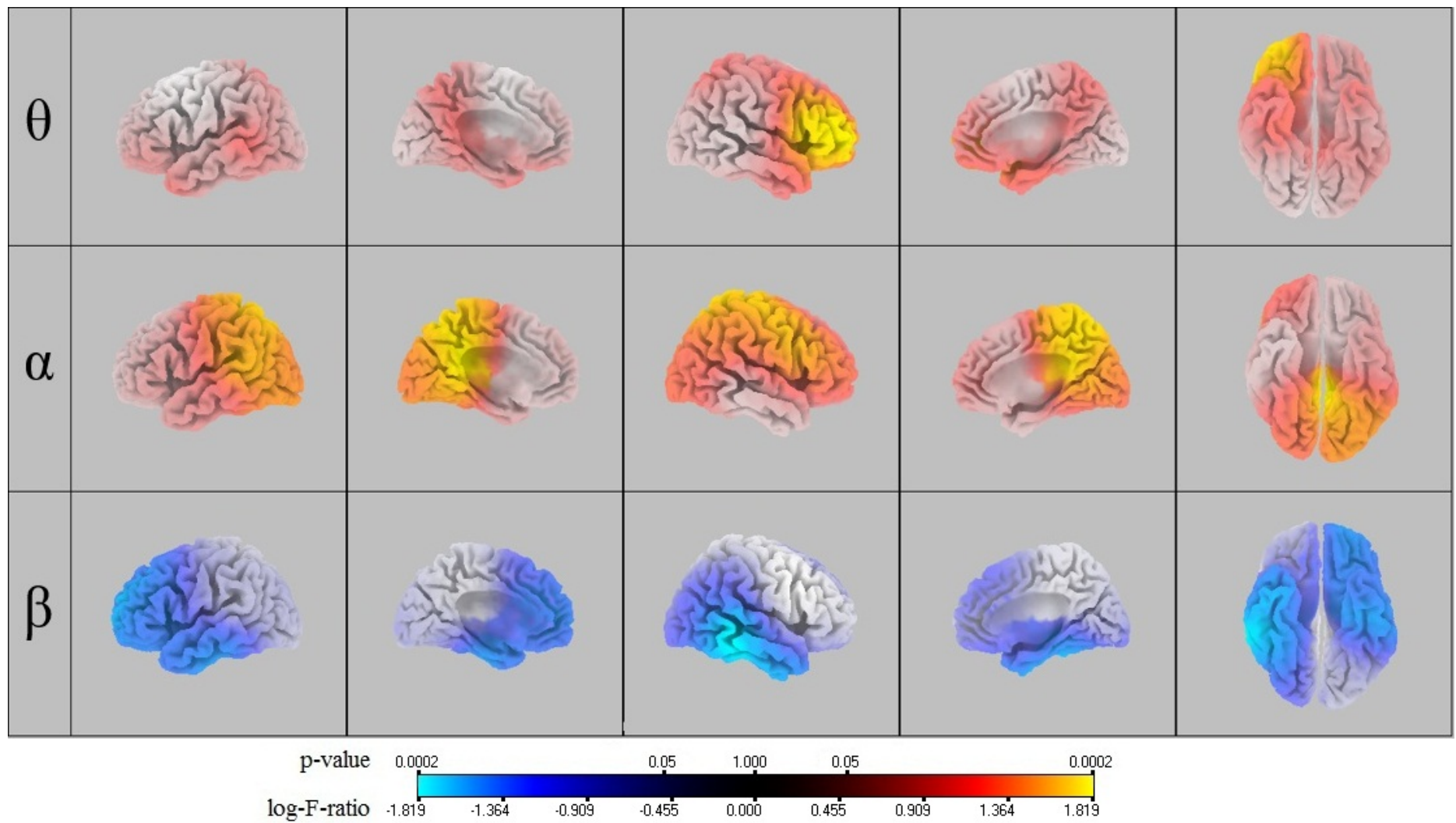

**Figure 3** – Between-group differences of cortical representation of the spectral density of main EEG rhythms after socialization. Red color is the higher activity in group of well-socialized gamers. Blue color is the higher activity in group of non-socialized gamers.